\documentclass[12pt]{article}
\usepackage{amssymb,amsmath}
\def \CC {\mathrm{Corr}}

\def \HH {\mathcal{H}}

\def \SS {\mathcal{S}}

\def \tr {\mathrm{Tr}}

\def \bs {\mathbf{s}}

\def \equals {\ = \ }
\def \plus {\ + \ }

\def \quasifree {\Gamma} 

\title{New measure of electron correlation}

\author{
Alex D.\  GOTTLIEB \footnote{ Wolfgang Pauli Institute,
Nordbergstr. 15, A--1090 Wien, Austria (alex@alexgottlieb.com).
} \\
and Norbert J.\ MAUSER \footnote{ WPI c/o Fak. f.\ Math., Univ.
Wien, Nordbergstr. 15, A--1090 Wien, (mauser@courant.nyu.edu). } }

\date{}

\begin{document}
\maketitle

\begin{abstract}
We propose to quantify the ``correlation" inherent in a
many-electron (or many-fermion) wavefunction $\psi$ by comparing
it to the unique uncorrelated state that has the same $1$-particle
density operator as does $|\psi\rangle \langle \psi |$.
\end{abstract}


\noindent

Electron correlation is of fundamental importance in quantum
chemistry and magnetism, but as yet there is no definitive way to
quantify it: given the wavefunction $\psi(x_1,x_2,\ldots,x_N)$
representing the state of a system of $N$ electrons, how much
``correlation" is there in that $N$-electron state? By definition,
a wavefunction that has the form of a Slater determinant
represents an ``uncorrelated" state, but how much ``correlation"
should be attributed to states  that are {\it not} represented by
a Slater determinant wavefunction?  Some measures of
``correlation" have already been advanced in the literature: the
``degree of correlation"
\cite{GrobeRzazewskiEberly,GrobeEberlyPRL} and the ``correlation
entropy"
\cite{Ziesche,EsquivelEtAl,GersdorfEtAl,ZiescheEtAl,ZiescheEtAl2},
for example. These correlation measures depend completely upon the
eigenvalues of the $1$-particle statistical operator $\gamma$,
i.e., the operator with the integral kernel
\begin{equation}
\label{gamma}
     \gamma(x,y) = N \int \psi(x,z_2,\ldots,z_N) \overline{\psi(y,z_2,\ldots,z_N) }
    \ dz_2 dz_3 \cdots dz_N
\end{equation}
(supposing that the wavefunction $\psi$ has norm $1$, so that the
trace of $\gamma$ equals the number of electrons). The ``degree of
correlation" is inversely proportional to the sum of the squares
of the eigenvalues of $\gamma$, and the ``correlation entropy" is
the Shannon entropy of those eigenvalues. Such measures ascribe
the same amount of ``correlation" to all wavefunctions that have
the same $1$-particle statistical operator.  We feel it imposes a
severe conceptual limitation to have to say that all wavefunctions
having the same $1$-particle operator (or the same $2$-particle
operator, for that matter) contain the same amount of correlation.
Here we propose a new measure of correlation which does not suffer
that limitation.

States represented by Slater determinant wavefunctions are the
only {\it pure} states are deemed ``uncorrelated"; but certain
{\it mixed states} should also be regarded as ``uncorrelated",
namely, the mixed ``quasifree states."  These have random particle
number and must be represented by density operators (i.e.,
statistical operators of trace $1$) on the fermion Fock space. Let
$\HH$ be the $1$-electron Hilbert space, and let $a_f^{\dagger}$
and $a_f$ denote the electron creation and annihilation operators
for $f \in \HH$.   A density operator $\quasifree$ on the fermion
Fock space over $\HH$ represents a ``quasifree state"
\cite{Quazzz} if
\begin{equation}
\label{Quasifree}
     \tr
     \big(
       \quasifree
       a_{f_1}^{\dagger} a_{f_2}^{\dagger} \cdots  a_{f_m}^{\dagger}a_{g_n}\cdots a_{g_2} a_{g_1}
     \big)
     \equals \delta_{mn}\ \det \big(
     \tr( \quasifree a^{\dagger}_{f_i} a_{g_j} )
     \big)_{i,j=1}^n
\end{equation}
for all $n$ and all $f_1,\ldots,f_m,g_1,\ldots,g_n \in \HH$
($\delta_{mn}$ denotes Kronecker's delta). The two-point
correlations $\tr( \quasifree a^{\dagger}_{f} a_{g} )$ determine
all higher correlations under the quasifree state $\quasifree $,
and in this sense a density operator satisfying (\ref{Quasifree})
may be called ``uncorrelated."   The $1$-particle operator
$\gamma$ defined by
\begin{equation}
\label{reducesTogamma}
     \langle g, \gamma f \rangle \equals \tr( \quasifree  a_f^{\dagger}a_g)
\end{equation}
has trace equal to the average particle number under $\quasifree$,
i.e., $\tr(\gamma) = \tr(\quasifree \mathrm{N})$ where
$\mathrm{N}$ denotes the number operator.

In case the average particle number is finite, we can reconstruct
$\quasifree$ from the $1$-particle operator $\gamma$ as follows.
Let $\{\phi_i\}_{i=1}^{\infty}$ be a complete system of
eigenvectors of $\gamma$ with
\begin{equation}
\label{NaturalOrbitals}
     \gamma (\phi_i) \equals \lambda_i \phi_i\ .
\end{equation}
The eigenvalues $\lambda_i$ are all between $0$ and $1$; we will
interpret them as ``occupation probabilities." Let $\SS$ denote
the class of all finite sets of positive integers, including the
empty set.  We can choose a random member of $\SS$, that is, we
can form a random set $\bs$ of positive integers, by including $i$
in $\bs$ with probability $\lambda_i$ (and excluding it with
probability $1 -\lambda_i$) independently of all other positive
integers which may or may not be included in $\bs$. This procedure
produces the set $\bs$ with probability
\begin{equation}
\label{p}
    p(\bs)
    \equals
    \prod_{i \in \bs} \lambda_i  \prod_{i \notin \bs} (1 - \lambda_i ) \ .
\end{equation}
For any nonempty $\bs = \{i_1,\ldots,i_n\} \in \SS$, let
$\Psi_{\bs}$ denote a normalized Slater determinant in the natural
orbitals $\phi_{i_1},\ldots, \phi_{i_n}$ (the phase of
$\Psi_{\bs}$ is arbitrary) and define
\[
    \mathrm{P}_{\bs} \equals
    0 \oplus \cdots \oplus 0 \ \oplus
    \stackrel{n-particle\ space}{\ |\Psi_{\bs}\rangle \langle \Psi_{\bs} |\ }\oplus \ 0
    \oplus 0 \oplus \cdots\
\]
--- a projector on the Fock space; in case $\bs$ is the empty set, let $\mathrm{P}_{\bs}$ denote
projection onto the vacuum space.   It can be shown
\cite{CanBeShown} that
\begin{equation}
\label{ExplicitQuasifree}
     \quasifree  \equals \sum_{\bs \in \SS} p(\bs) \mathrm{P}_{\bs}
\end{equation}
satisfies (\ref{Quasifree}) and (\ref{reducesTogamma}). Slater
determinants are special cases of quasifree states: if $\Psi$ is a
Slater determinant in the $1$-electron orbitals
$\psi_1,\ldots,\psi_N$ then the projector
\[
    \mathrm{P}_{\bs} \equals
    0 \oplus \cdots \oplus 0 \ \oplus
    \stackrel{N-particle\ space}{\ |\Psi \rangle \langle \Psi |\ }
    \oplus \ 0 \oplus 0 \oplus \cdots\
\]
represents a quasifree state whose $1$-particle statistical
operator is the projector onto the span of
$\{\psi_1,\ldots,\psi_N\}$.

Our idea is to quantify the ``correlation" in a many-electron
wavefunction $\psi$ by comparing it to the unique uncorrelated
(quasifree) state $\quasifree$ having the same $1$-particle
statistical operator $\gamma$ as does $|\psi\rangle \langle \psi
|$. The more $|\psi\rangle\langle \psi |$ resembles $ \quasifree
$, the less ``correlated" we consider $\psi$ to be: we would like
to quantify the correlation inherent in $\psi$ by some measure of
the dissimilarity of $|\psi\rangle\langle \psi |$ and $ \quasifree
$. There are several ways one might measure this ``dissimilarity"
and thereby quantify electron correlation; our choice is the
following one: we identify $\psi$ with an $N$-particle vector in
the Fock space and define
\begin{equation}
\label{correlation}
    \CC(\psi) \equals -  \log\  \langle \psi,  \quasifree  \psi \rangle\ ,
\end{equation}
where $\quasifree$ is the quasifree density (\ref{ExplicitQuasifree}) determined by the
$1$-particle statistical operator (\ref{gamma}) for $\psi$.   This quantity is
nonnegative, and it equals $0$ if and only if $\psi$ is a Slater
determinant. It can be seen from the discussion in the following
paragraph that $\CC(\psi) < \infty$.

It is easy to calculate $\CC(\psi)$ if the expansion of $\psi$ in
Slater determinants in the eigenvectors of the $1$-particle
statistical operator (\ref{gamma}) is available. The eigenvectors
of $\gamma$ are called natural orbitals and the corresponding
eigenvalues lie between $0$ and $1$ \cite{Loewdin}. Here,
``natural orbitals" will refer only to eigenvectors of $\gamma$
with {\it nonzero} eigenvalues. Recalling the notation in formulas
(\ref{NaturalOrbitals}) and (\ref{p}) above, let $\SS_N$ consist
of all sets of $N$ positive integers, each of which is the index
of a natural orbital. The wavefunction $\psi$ is a superposition
of Slater determinants in the natural orbitals, even when the
natural orbitals don't span the whole $1$-particle space
\cite{StrongerColeman}. Thus $
    \psi = \sum_{\bs \in \SS_N} c(\bs) \Psi_{\bs}\ ,
$ where $\Psi_{\bs}$ denotes a Slater determinant in the orbitals
indexed by $\bs$, and we have
\begin{equation}
\label{correlationRecipe}
   \CC(\psi) \equals - \log\ \sum_{\bs \in \SS_N} p(\bs) |c(\bs)|^2\ ,
\end{equation}
where $p(\bs)$ is as defined in (\ref{p}).

\noindent {\bf Remark 1: a formula for $\CC(\psi)$ in $2$-particle
systems.} \qquad

When $\psi(x,y) = -\psi(y,x)$ is a $2$-electron wavefunction
 $\CC(\psi)$ is a functional
of the eigenvalues of the $1$-particle statistical operator.  In
this case, it is known \cite{Footnote} that there exist an
orthonormal system $ \{ f_1,\ g_1,\ f_2,\ g_2,\ \ldots \ \}  $ of
$1$-electron orbitals and nonnegative numbers $p_1,p_2,\ldots$
such that
\begin{equation}
\label{specialSchmidt} \psi(x,y) \equals
     \sum_{j = 1,2,\ldots}
     \sqrt{p_j}
     \tfrac{1}{\sqrt{2}}
         \big(
        f_j(x) g_j(y)
        -
        g_j(x) f_j(y)
        \big)\ .
\end{equation}
The wavefunctions $f_i$ and $g_i$ are natural orbitals with
occupation probabilities $p_i$, for
\[
    \gamma \equals \sum_i p_i \big( |f_i \rangle \langle f_i | + |g_i \rangle \langle g_i | \big)\ .
\]
In this case, one can calculate that
\begin{equation}
\label{2-particleFormula}
     \CC(\psi)
    \equals
     - \log\  \sum_i p_i \Big\{ p_i
    \prod_{j:j\ne i} (1-p_j)\Big\}^2
\end{equation}
using formula (\ref{correlationRecipe}). Formula
(\ref{2-particleFormula}) shows that $\CC(\psi)$ can be
arbitrarily large, for the $2$-electron wavefunction
(\ref{specialSchmidt}) can have arbitrarily small coefficients
$p_j$ \cite{Fermion}.

It is interesting to apply formula (\ref{2-particleFormula}) to an
exactly solvable model: the ``two-site Hubbard model" that was
investigated in \cite{ZiescheEtAl} as a study of ``correlation
entropy."  This simple model can be used to illustrate the
``kinetic exchange" correction to the Heitler-London theory of
diatomic hydrogen \cite{Fazekas}. There are two fixed spatial
orbitals, labelled $1$ and $2$, whose span must accomodate two
electrons, and the Hamiltonian is
\[
    H \equals -t\left(
    a^{\dagger}_{1\uparrow} a_{2\uparrow} + a^{\dagger}_{2\uparrow} a_{1\uparrow}
    +
    a^{\dagger}_{1\downarrow} a_{2\downarrow} + a^{\dagger}_{2\downarrow} a_{1\downarrow}
    \right)
    \ + \ U\left(
    a^{\dagger}_{1\uparrow} a_{1\uparrow} a^{\dagger}_{1\downarrow} a_{1\downarrow}
    +
    a^{\dagger}_{2\uparrow} a_{2\uparrow} a^{\dagger}_{2\downarrow} a_{2\downarrow}
    \right)\ ,
\]
where $U$ is the on-site repulsion energy (or attraction energy if
$U < 0$).   The $2$-electron ground state of $H$ depends only upon
the dimensionless interaction parameter $u=U/t$.  Denote this
ground state by $\psi_u$.  The ground state $\psi_0$ for $u=0$ is
a Slater determinant, and $\psi_u$ tends towards the maximally
correlated Heitler-London state $\tfrac{1}{\sqrt{2}} \big(
a^{\dagger}_{1\uparrow} a^{\dagger}_{2\downarrow} +
a^{\dagger}_{2\uparrow} a^{\dagger}_{1\downarrow} \big) |\ \rangle
$ as $u$ tends to $+ \infty$.   The correlation entropy (i.e., the
Shannon entropy of the spectrum of the $1$-particle operator) of
$\psi_u$ was found \cite{ZiescheEtAl} to be strictly increasing in
$|u|$, approaching its maximum possible value as
$|u|\longrightarrow \infty$.  The same is true of $\CC(\psi_u)$.
However, $\CC(\psi_u)$ seems better behaved than the correlation
entropy $S(\psi_u)$ for small values of $|u|$:  the former is
infinitely differentiable at $u=0$ but the latter is only
differentiable once there, and $\lim\limits_{u \rightarrow 0}
S(\psi_u)/\CC(\psi_u) = \infty$.  If $S(\psi_u)$ is normalized so
that $\lim\limits_{u \rightarrow \infty} S(\psi_u)/\CC(\psi_u) =
1$, then $S(\psi_u)$ is larger than $\CC(\psi_u)$ for all $|u|>0$.

\noindent {\bf Remark 2: $\CC(\psi)$ is not a function of the
$1$-particle operator.} \qquad

We have seen that $\CC(\psi)$ is a function of the spectrum of the
$1$-particle statistical operator in case $\psi$ is a $2$-electron
wavefunction.  In general, however, $\CC(\psi)$ is not a function
of the $1$-particle statistical operator.   To show this, we
exhibit two $3$-particle wavefunctions that have the same
$1$-particle statistical operator but contain different amounts of
correlation. Let $e_1,e_2,\ldots,e_6$ be six orthonormal
$1$-electron orbitals, and let us denote $3$-particle Slater
determinants in these orbitals by listing the three indices
involved (in increasing order) in between vertical lines, so that,
for example,
\[
 | 245 | \equals
     \frac{1}{\sqrt{6}}
     \big(
      | e_2 e_4 e_5 \rangle + |e_5  e_2  e_4 \rangle + |e_4  e_5  e_2 \rangle
      - |e_4  e_2  e_5 \rangle - |e_2  e_5  e_4 \rangle - |e_5  e_4  e_2
      \rangle \big)\ .
\]
Consider the two $3$-particle wavefunctions
\begin{eqnarray*}
\Psi & = &  \sqrt{\tfrac23}\ |135| \plus \sqrt{\tfrac13}\ |246|
\\
 \Phi & = & \sqrt{\tfrac13}\ \Big( |123| + |345| +|156| \Big)\ .
\end{eqnarray*}
These have the same $1$-particle statistical operator
\[
        \left[
                    \begin{array}{cccccc}
                    2/3 & 0 &0 & 0&0& 0\\
                    0 & 1/3 & 0 & 0& 0& 0\\
                     0 & 0 & 2/3 & 0 & 0& 0\\
                    0  & 0 & 0 & 1/3 & 0 & 0\\
                    0  & 0 & 0 & 0 & 2/3 & 0 \\
                     0 & 0 & 0 & 0 &    0 & 1/3
                    \end{array}
        \right]
\]
but $\CC(\Psi) \thickapprox 4.08$  and $ \CC(\Phi) \thickapprox
5.51$, taking logarithms to the base $2$ in formula
(\ref{correlationRecipe}). (The matrix above displays only the
matrix elements $\langle e_i| \gamma e_j \rangle$; outside this
finite block all matrix elements are $0$.)

Similar examples can be contrived to show that $\CC$ is not a
function of the $m$-particle statistical operator for any $m>1$ either \cite{however}.

\noindent {\bf Remark 3: $\CC$ for mixed states.} \qquad

The measure (\ref{correlation}) of
fermion correlation can be extended to from pure states (given by wavefunctions)
to mixed states (given by density operators).  If $D$ is a density operator, we would require
``the correlation in $D$" to equal $0$ if and only if $D$ is a quasifree density,
and of course ``the correlation in $D$" must equal $\CC(\psi)$ in case $D=|\psi\rangle \langle \psi |$.
These requirements are met by
\begin{equation}
\label{ForMixedStates}
 Corr(D) = - 2\ \log \big( \tr (D^{1/2}
\quasifree D^{1/2} )^{1/2} \big) \ ,
\end{equation}
where $\quasifree$ is the quasifree density (\ref{Quasifree})
corresponding to the $1$-particle statistical operator $\gamma$
defined by $\langle g, \gamma f \rangle = \tr( D  a_f^{\dagger}
a_g)$ \cite{Supposing}. Formula (\ref{ForMixedStates})  uses the
generalized ``transition probability" \cite{Uhlmann} or
``fidelity" \cite{Jozsa} between $D$ and $ \quasifree $:
\[
   \tr (D^{1/2} \quasifree D^{1/2} )^{1/2}
   \equals \big\| D^{1/2}\quasifree^{1/2} \ \big\|_{trace}
   \equals \tr ( \quasifree^{1/2} D \quasifree^{1/2} )^{1/2}\ ,
\]
This generalized transition probability recommends itself as a
measure of the closeness of any two density operators --- $D$ and
$ \quasifree $ in this case --- because it enjoys several
properties that distinguish it as a useful quantity in
quantum information geometry \cite{Bures,Uhlmann,Uhlmann2,PetzSudar}.

Note that (\ref{ForMixedStates}) assigns positive correlation to most mixtures of uncorrelated states.
For example, if $\phi_1$ and $\phi_2$ are orthogonal $1$-particle wavefunctions, then
\[
   Corr \big( \tfrac12 |\phi_1 \rangle \langle \phi_1 |+\tfrac12 |\phi_2 \rangle \langle \phi_2 | \big) \equals 1
\]
(taking logarithms to the base $2$).  It may seem odd to assign positive correlation to a $1$-particle state,
but it is due to our identification of ``uncorrelated" with ``quasifree."  This means that a
state is deemed ``uncorrelated" only if the occupations of its natural
orbitals are statistically independent.  In the example, the natural orbitals $\phi_1$ and $\phi_2$ are
each occupied with probability $1/2$, but the probability that {\it both} natural orbitals
are occupied (at the same time) is not
$1/2 \times 1/2$.  We consider the state to be ``correlated" because the occupation numbers
of its natural orbitals are correlated random variables.

\noindent {\bf Conclusion}\qquad

We have proposed a new way to measure the ``correlation" in states of
many-electron (or other many-fermion) systems. The idea is to compare the state to its
unique reference state: the quasifree state with the same
$1$-particle statistical operator. To quantify the agreement
between the many-electron state and its reference state, we have
chosen the negative logarithm of the overlap, viz. formulas
(\ref{correlation}) and (\ref{ForMixedStates}).

Our correlation measure, as formulated above, applies only to {\it
finite} systems (atoms and molecules) but not infinite systems
(solids).  Other interesting measures of correlation are available
for (infinite) uniform electron gases \cite{Ziesche,GoriGiorgiZiesche,Ziesche2000}.  Can a serviceable
formulation of our correlation measure be found for infinite
systems as well?

The most intriguing property of the measure $\CC(\psi)$ is the
fact that it is not a function of the $1$-particle statistical
operator: wavefunctions with the same $1$-particle statistical
operator may have different amounts of ``correlation."  For
wavefunctions of $2$-electron systems, however, $\CC(\psi)$ is a
function of the eigenvalues of the $1$-particle operator, viz.
formula (\ref{2-particleFormula}). Further investigation of the
comportment of $\CC(\psi)$ is underway, to see whether it conforms
to what we would intuitively expect from a measure of electron
correlation.

\bigskip

\noindent {\bf Acknowledgements.} {\it This work was supported by
the Austrian Ministry of Science (BM:BWK) via its grant for the
Wolfgang Pauli Institute and by the Austrian Science Foundation
(FWF) via the START Project (Y-137-TEC) of N. Mauser, and also by
the European network HYKE
funded by the EC as contract HPRN-CT-2002-00282. \\
We are grateful to Armin Scrinzi and Claude Bardos \cite{BGGM,BGGM2} for valuable
discussions.}



\begin{thebibliography}{X}
\bibitem{Fermion}  Note that this argument does not work for fermions with a
finite-dimensional $1$-particle space: in such cases the
correlation of a $2$-fermion wavefunction has a finite upper
bound.
\bibitem{Supposing} Here $Corr(D)$ is only defined for $D$ such that $\tr(D
\mathrm{N})< \infty$, i.e., for mixed states with finite average
particle number, because we want $\quasifree$ in
(\ref{ForMixedStates}) to be a density operator.
\end{thebibliography}
\end{document}